\documentclass[1p]{elsarticle}
\usepackage{amssymb}
\usepackage{amsmath}
\usepackage[dvips]{color}

\begin{document}
\begin{frontmatter}

\title{Is the nonmonotonic behavior in the cross section of $\phi$ photoproduction
near threshold a signature of a resonance?}

\author[ntuphys,cts]{Alvin Kiswandhi}
\author[ntuphys]{Ju-Jun Xie}
\author[ntuphys,cts]{Shin Nan Yang}

\address[ntuphys]{Department of Physics, National Taiwan University,
Taipei 10617, Taiwan}
\address[cts]{Center for Theoretical Sciences, National Taiwan University,
Taipei 10617, Taiwan}

\begin{abstract}

We  study whether the nonmonotonic behavior found in  the
differential cross section of the $\phi$-meson photoproduction near
threshold can be described by a resonance. Namely, we add a
resonance to a model consisting of Pomeron  and $(\pi,
\eta)$ exchange by fiat and see if, with a suitable assignment of spin
and parity, mass and width, as well as the coupling constants, one
would be able to obtain a good description to all the data reported
by the LEPS collaboration  in the low-energy region.  The resonant
contribution is evaluated by using an effective Lagrangian approach. We find
that, with the assumption of a $J^P=3/2^-$ resonance with  mass of
$2.10\pm 0.03$ GeV and width of $ 0.465\pm 0.141$ GeV, LEPS data can
indeed be well described. The ratio of the helicity amplitudes
$A_{\frac 12}/A_{\frac 32}$ calculated from the resulting coupling
constants differs in sign from that of the known $D_{13}(2080)$. We
further find that the addition of this postulated resonance can
substantially improve the agreement between the existing theoretical
predictions and the recent $\omega$ photoproduction data if  a large
value of the OZI evading parameter $x_{\textrm{OZI}}=12$ is assumed for the
resonance.
\end{abstract}

\begin{keyword}
Photoproduction, $\phi$ meson, nucleon resonance, Pomeron
\PACS 13.60.Le \sep 25.20.Lj \sep 14.20.Gk
\end{keyword}

\end{frontmatter}

A well-established feature in the $\phi$-meson photoproduction
reaction at high energies is that it is dominated by the diffractive processes,
which are conveniently described by the $t$-channel Pomeron $(P)$
exchange \cite{bauer78,donnachie87}. In the low-energy region, the
nondiffractive processes of the pseudoscalar $(\pi, \eta)$-meson
exchange are known to contribute  \cite{bauer78}. In addition, many
other processes, including nucleon and nucleon-resonance exchanges,
second Pomeron exchange, $t$-channel scalar meson and glueball
exchanges, and $s\bar s$-cluster knockout have also been extensively
studied \cite{titov97,titov_polarization,williams98,titov99,oh01,titov03,titov07}.
However, no definite conclusion has been reached because of the
limited experimental data.

Recently, a local maximum in the differential cross sections
of $\phi$ photoproduction on  protons at  forward angles  at around
$E_{\gamma}\sim 2.0$ GeV, has been observed by the LEPS
collaboration \cite{leps05}. Models which consist of $t$-channel
exchanges \cite{titov97,titov_polarization,williams98,titov99,oh01,titov03,titov07}
have not been able to account  for such a nonmonotonic behavior.

Typically, local maxima in the cross sections are often associated
with resonances. Effects of the resonances in $s$- and $u$-channels
up to mass 2 GeV have been investigated in Refs.
\cite{titov07,zhao99}. Ref. \cite{zhao99} used a constituent quark
model with $\textrm{SU}(6)\otimes \textrm{O}(3)$ symmetry and
included explicitly excited resonances with quantum numbers $n\leq
2$, while Ref. \cite{titov07} considered all the known 12 resonances
below 2 GeV listed in Particle Data Group \cite{PDG02}, with
coupling constants determined by the available data
\cite{besch74,anciant00} at large momentum transfers. The resonances
are found to play non-negligible role, especially in polarization
observables. However, no local maximum as observed in Ref. \cite{leps05}
was obtained.

In this Letter, we  study whether the nonmonotonic behavior found in
Ref. \cite{leps05} can be described by a resonance. Namely, we will add a
resonance to a model consisting of Pomeron   and $(\pi,
\eta)$ exchange by fiat and see if, with a suitable assignment of spin
and parity, mass and width, as well as the coupling constants, one
would be able to obtain a good description of all the data reported
by the LEPS collaboration, which include the angular and energy
dependence of the differential cross section and decay angular
distributions in the Gottfried-Jackson frame, in the low-energy
region from threshold to $E_\gamma = 2.37$ GeV. Since the local
maximum appears quite close to the threshold, we will investigate,
as a first step, the possibility of the spin of the resonance being
either $1/2$ or $3/2$. Similar analysis was carried out in a
coupled-channel model \cite{ozaki09}. However, the analysis was
marred by a confusion in the phase of the Pomeron-exchange amplitude
\cite{hosaka10}.
\begin{figure}
\includegraphics[width = 1.0\linewidth,angle=0]{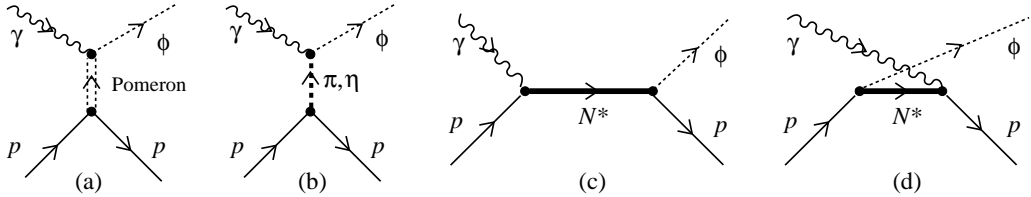}
\caption{ (a) Pomeron-exchange, (b) $(\pi,\eta)$-exchange, and (c)
$s$- and $u$-channel $N^*$-exchange diagrams for $\gamma p \to \phi
p$ reaction.} \label{gammapdiagram}
\end{figure}

We first define the kinematical variables $k$, $p_i$, $q$, and $p_f$
as the four-momenta of the incoming photon, initial proton, outgoing
$\phi$-meson, and final proton, respectively; and $s=(k+p_i)^2 $,
$t=(q-k)^2 $, and $u=(p_f-k)^2$. The full amplitude in our model
consists of Pomeron-exchange, $t$-channel $(\pi, \eta)$-exchange,
and the $s$- and $u$-channel $N^*$-exchange amplitudes.
The Pomeron-exchange amplitude can be expressed as \cite{titov03,titov07},
\begin{eqnarray}
{\cal M}_P &=& -\bar{u}(p_f,\lambda_{N'})M(s,t)\Gamma^{\mu \nu} u(p_i,\lambda_N) \nonumber \\
&\times&\varepsilon^*_{\mu}(q,\lambda_{\phi})\varepsilon_{\nu}(k,\lambda_{\gamma}),
\label{ampa}
\end{eqnarray}
where $\varepsilon_{\mu}(q,\lambda_{\phi})$ and
$\varepsilon_{\nu}(k,\lambda_{\gamma})$ are the polarization vectors
of the $\phi$-meson and photon with helicities $\lambda_{\phi}$ and
$\lambda_{\gamma}$, respectively; and $u(p ,\lambda_N)$  the Dirac
spinor of the nucleon with momentum $p $   and helicity $\lambda_N$.
The explicit form for the transition
operator $\Gamma^{\mu \nu}$ can be found in Refs. \cite{titov03,titov07}
and the scalar function $M(s,t)$  is given by the Reggeon
parametrization,
\begin{eqnarray}
M(s,t) &=& C_P
F_1(t)F_2(t)\frac{1}{s}\left(\frac{s-s_{th}}{4}\right)^{\alpha_P(t)}\nonumber \\
&\times&\text{exp}\left[{-i\pi\alpha_P(t)/2}\right],
\end{eqnarray}
where we introduce an additional threshold factor $s_{th}$ as also
done in Refs.~\cite{williams98,titov99,titov03} to adjust the shape
of the energy dependence of the Pomeron amplitude near the threshold.
Also, $F_1(t)$ and $F_2(t)$ are the isoscalar electromagnetic form factor
of the nucleon and the form factor of $\gamma\phi P$ coupling,
respectively, and are taken to be of the form given in Refs. 
\cite{donnachie87,titov03}. As in \cite{titov03}, we take
$\alpha_P(t) = 1.08 + 0.25 t$, $\mu^2_0 = 1.1$ $\textrm{GeV}^2$, and
 $C_P = 3.65$ which is obtained by fitting to the total
cross sections data at high energy. We choose $s_{th} = 1.3$
$\textrm{GeV}^2$ by matching the forward differential cross sections data at around
$E_\gamma = 6$ $\textrm{GeV}$ \cite{durham}.

  The contribution of the $t$-channel $(\pi, \eta)$ exchange to the $\phi$ photoproduction
is rather well understood. We  use the same set of parameters for
the pseudoscalar-exchange amplitude as adopted in Ref.
\cite{titov07} except, in their notation,   $g_{\eta NN} =
1.12$~\cite{wentai} and $\Lambda_{\pi(\eta)} = 1.2$ GeV, the cutoff
in the form factor.

   We will consider the cases where the spin of the resonance is either
$1/2$ or $3/2$. The interaction Lagrangian densities which describe
the coupling of spin-$1/2$ and $3/2$ particles to $\gamma N$ and
$\phi N$, can in general be written as
~\cite{wentai,pascalutsa,feu},
\begin{eqnarray}
{\cal L}_{\phi N N^*}^{1/2^\pm} &=& g_{\phi N N^*}^{(1)} \bar{\psi}_{N} \Gamma^\pm \gamma^\mu \psi_{N^*} \phi_{\mu} \nonumber \\
&+& g_{\phi N N^*}^{(2)}\bar{\psi}_{N} \Gamma^\pm \sigma_{\mu \nu}
F^{\mu \nu} \psi_{N^*},  \label{phiNNstaronehalf}\\
{\cal L}_{\phi N N^*}^{3/2^\pm} &=& ig^{(1)}_{\phi N N^*}\bar{\psi}_{N} \Gamma^\pm \left(\partial^\mu\psi_{N^*}^\nu\right) \tilde{G}_{\mu \nu} \nonumber \\
&+& g^{(2)}_{\phi N N^*} \bar{\psi}_{N} \Gamma^\pm \gamma^5 \left(\partial^{\mu} \psi_{N^*}^\nu \right)G_{\mu \nu} \nonumber \\
&+& ig^{(3)}_{\phi N N^*} \bar{\psi}_{N} \Gamma^\pm \gamma^5  \gamma_\alpha \nonumber \\
&\times&\left(\partial^\alpha \psi^\nu_{N^*} - \partial^\nu \psi^\alpha_{N^*}\right)
\left(\partial^\mu G_{\mu\nu}\right), \label{phiNNstar}
\end{eqnarray}
where $G_{\mu \nu}= \partial_{\mu}\phi_{\nu} -
\partial_{\nu}\phi_{\mu}$ represents the $\phi$-meson field tensor and
$\tilde{G}_{\mu \nu} = {1 \over 2} \epsilon_{\mu\nu\alpha\beta}
G^{\alpha\beta}$ with $\epsilon^{0123} = +1$. The operator
$\Gamma^\pm$ are given by $\Gamma^+=1$ and $\Gamma^-=\gamma_5$,
depending on the parity of the resonance $N^*$. For the $\gamma N
N^*$ vertices, one simply changes $g_{\phi N N^*} \to e g_{\gamma N
N^*}$ and $\phi_\mu \to A_\mu$. However,   current conservation
consideration fixes $g^{(1)}_{\gamma N N^*}$ for $J^P = 1/2^\pm$
resonances to be zero.  In addition, the term proportional to
$g^{(3)}_{\gamma N N^*}$ in  the Lagrangian densities of Eq.
(\ref{phiNNstar}) vanishes in the case of real photon. The
form factor for the vertices used in the $s$- and $u$-channel
diagrams, $F_{N^*}(p^2)$, is taken
 as    $F_{N^*}(p^2)=
\Lambda^{4}/[\Lambda^{4} + (p^2-M^2_{N^*})^2]$ \cite{Mosel,
feuster}, with $\Lambda$ the cutoff parameter for the virtual
$N^*$. We choose $\Lambda = 1.2$ GeV for all resonances. The effect
of the width is taken into account in a Breit-Wigner form by
replacing the usual denominator $p^2-M^2_{N^*} \to
p^2-M^2_{N^*}+iM_{N^*}\Gamma_{N^*}$, with $\Gamma_{N^*}$ is the
total decay width of $N^*$. Since $u<0$, we take $\Gamma_{N^*} = 0 $
MeV for the $u$-channel propagator.

With the interaction Lagrangian densities given in Eqs.
(\ref{phiNNstaronehalf},\ref{phiNNstar}), it is straightforward to
write down the invariant amplitudes of the  $s$- and $u$-channel
exchange diagrams of the corresponding $N^*$. In tree-level approximation,
only the products like $eg_{\gamma N N^*}g_{\phi N N^*}$, enter in
the invariant amplitudes. They are determined with the use of MINUIT,
by fitting to the experimental data \cite{leps05}, including differential cross section at forward angle as a function of photon energy and
differential cross section as a function of $t$ at different photon energies, as well as to
five decay angular distributions at two photon energies.

We find that with the assignments of $J^P = 1/2^\pm$ to the
resonance, it is not possible to produce  the nonmonotonic behavior
near threshold, in contrast to the finding of Refs.
\cite{ozaki09,hosaka10}.

\begin{figure}[t]
\begin{center}
\includegraphics[width=0.5\linewidth,angle=0]{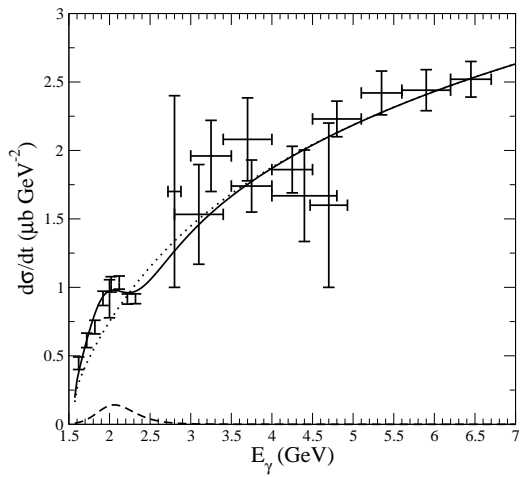}
\end{center}

\caption{ Differential cross section of $\gamma p \to \phi p$ at
forward direction as a function of photon energy $E_\gamma$. The
dotted, dashed, and solid lines denote contributions from
nonresonant, resonance with $J^P = 3/2^-$, and their sum,
respectively. Data are from Refs.~\cite{leps05,durham}. }
\label{D13_1}
\end{figure}

For the assignments of $J^P = 3/2^\pm$, we find that both parities can
describe the differential cross section at forward angle well and can also describe other observables with
comparable quality. The resulting $\chi^2/N$, and (mass, width)  in
unit of GeV, for the case of $3/2^+$ and $3/2^-$ are 1.066 and
$(2.05\pm 0.06,0.450\pm 0.111)$, and 0.983 and $(2.10\pm 0.03,
0.465\pm 0.141)$, respectively. This leads us to the problem of
determining the parity of the resonance.

To resolve this question, we perform a stability check against
changes in Pomeron contribution, whose low-energy behavior is not
yet fully understood. It turns out that the extracted properties of the
resonances are more sensitive with respect to the variation in the
Pomeron parameters if the positive parity is chosen. Therefore,  we
prefer the choice of $J^P = 3/2^-$.   The coupling constants and the
extracted mass and width of the $J^P = 3/2^-$ resonance  are given
in Table \ref{tab:nstar-}.

\begin{figure*}[htbp]
\begin{center}
\vspace{0.5cm}
\includegraphics[width=1.0\linewidth,angle=0]{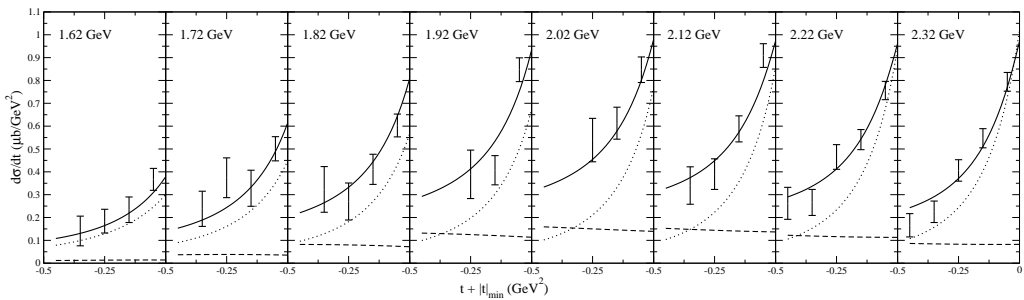}

\end{center}

\caption{{Differential cross sections of $\gamma p \to \phi p$ as a function of $t$
at eight different photon LAB energies. Data is taken from Ref.~\cite{leps05}. The notation is the same as in Fig.~\ref{D13_1}.}} \label{D13_2}
\end{figure*}

\begin{figure}[t]
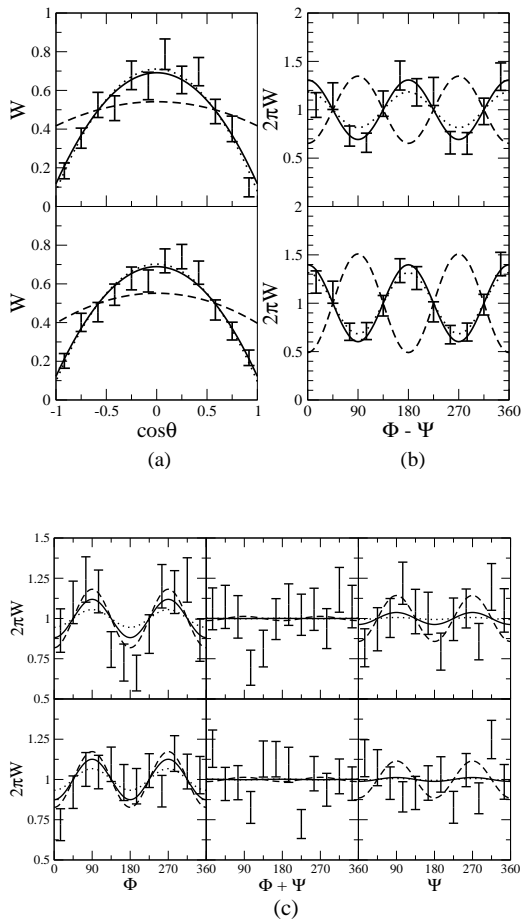

\begin{center}
\includegraphics[width=0.5\linewidth,angle=0]{figure4ab.eps}
\vspace{0.75cm}

\includegraphics[width=0.5\linewidth,angle=0]{figure4c.eps}
\vspace{-0.7cm}
\end{center}

\caption{{Our results obtained with  $J^P = 3/2^-$ resonance:
(a) decay angular distributions $ W(\cos \theta)$
(b) $W(\Phi - \Psi)$, and (c) $W(\Phi)$,
 $W(\Phi + \Psi)$, and $W(\Psi)$. All the decay angular
 distributions are given in two photon LAB energies,
 $1.97 - 2.17$ GeV (upper panel) and $2.17 - 2.37$ GeV (lower panel).
 Data is taken from Ref.~\cite{leps05}. The notation is the same as in Fig.~\ref{D13_1}.}}
 \label{D13_3}
\end{figure}

Our best fits with the choice of $J^P = 3/2^-$ to the experimental
energy dependence of the differential cross section at forward angle and angular dependence of the differential cross section
\cite{leps05,durham} are shown in Figs.~\ref{D13_1} and \ref{D13_2}, respectively. The dotted, dashed,
and solid lines correspond to contributions from nonresonant, \textit{i.e.},
Pomeron plus $(\pi, \eta)$-exchange, resonant, and the  full
results, respectively.  We find that no matter how the Pomeron
parameters are varied,  it is not possible to describe the
nonmonotonic behavior of the differential cross section at forward direction as a function
of photon energy with only the nonresonant contribution. One also
sees from Fig.~\ref{D13_2} that the addition of a resonance
markedly improves the agreement with the data on angular dependence.
\begin{table}
\caption{{The results for $N^*$ parameters with $J^P = 3/2^-$.}}

\begin{center}
\begin{tabular}{|c|c|}
\hline
$M_{N^*}$(GeV) & 2.10 $\pm$ 0.03 \\
$\Gamma_{N^*}$(GeV) &0.465 $\pm$  0.141  \\
\hline
$eg_{\gamma N N^*}^{(1)}g_{\phi N N^*}^{(1)}$ &$-0.186$ $\pm$  0.079 \\

$eg_{\gamma N N^*}^{(1)}g_{\phi N N^*}^{(2)}$ & $-0.015$ $\pm$  0.030 \\

$eg_{\gamma N N^*}^{(1)}g_{\phi N N^*}^{(3)}$ & $-0.02$ $\pm$  0.032 \\

$eg_{\gamma N N^*}^{(2)}g_{\phi N N^*}^{(1)}$ & $-0.212$ $\pm$  0.076 \\

$eg_{\gamma N N^*}^{(2)}g_{\phi N N^*}^{(2)}$ &$-0.017$ $\pm$  0.035 \\

$eg_{\gamma N N^*}^{(2)}g_{\phi N N^*}^{(3)}$ &$-0.025$ $\pm$  0.037 \\

\hline
\end{tabular}
\end{center} \label{tab:nstar-}
\end{table}

Our results for the decay angular distributions of the $\phi$-meson
in its rest frame (or the Gottfried-Jackson system, hereafter,
called GJ-frame), which can be expressed in terms of the
spin-density matrix elements
$\rho^{\alpha}_{ij}$~\cite{titov03,schillingnpb15397}, are shown in
Fig. \ref{D13_3}, where the contributions from nonresonant,  resonant,
and the full results are again denoted by dotted, dashed, and solid
lines, respectively. We   see that the data in $W(\cos \theta)$ at
both energies of $E_\gamma= 1.97-2.17$ GeV and $2.17-2.37$ GeV,
$W(\Phi-\Psi)$  at $2.17-2.37$ GeV, and $W(\Phi)$ again at
$2.17-2.37$ GeV can already be described relatively well by the
nonresonant contribution only and do not need strong modification
from a resonance. However, the rest of the distributions show some
discrepancies between nonresonant contribution and experimental data
and the inclusion of resonant contribution does help to reduce the
discrepancies. This is especially true for $W(\Phi-\Psi)$  at
$2.17-2.37$ GeV and $W(\Phi)$ at $1.97-2.17$ GeV, where the
nonresonant contribution does not describe satisfactorily the
experimental data. For both $W(\Phi+\Psi)$ and $W(\Psi)$, our model
still fail to give adequate agreement with the data which are of
rather poor quality with large error bars.

One might be tempted to identify the $3/2^-$ as the $D_{13}(2080)$
as listed in PDG \cite{PDG02}. The coupling constants given Table
\ref{tab:nstar-} can be used to calculate the ratio  of the helicity
amplitudes $A_{1/2}$ and $A_{3/2}$, though not their magnitudes
since only the product of the coupling constants for $\gamma NN^*$
and $\phi NN^*$ are determined. We obtain a value of
{$A_{1 \over 2}/A_{3 \over 2}=1.16$}, while it is
$-1.18$ for $D_{13}(2080)$. {Even though their
magnitudes are similar, the  relative sign  is, however, different.}
{For $J^P = 3/2^+$, we find that the value of $A_{1
\over 2}/A_{3 \over 2}=0.69$, again with positive sign.}

\begin{figure}[htbp]
\begin{center}
\vspace{0.5cm}

\includegraphics[width=0.5\linewidth,angle=0]{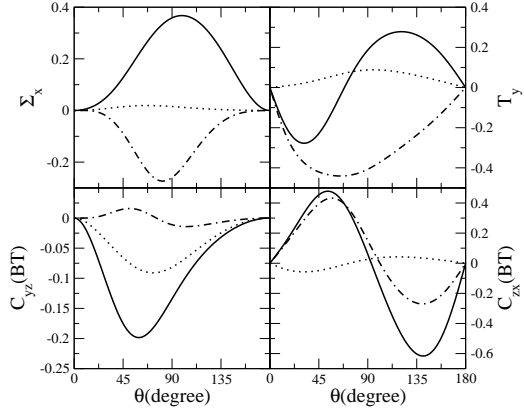}

\caption{{Single and double polarization observables
$\Sigma_x$, $T_y$, $C_{yz}^{BT}$, and $C_{zx}^{BT}$ taken at photon
laboratory energy $E_\gamma = 2$ GeV. The solid and dash-dotted
lines correspond to our results with the choices of $J^P = 3/2^-$ and
$J^P = 3/2^+$, respectively, while the dotted lines denote the
nonresonant contribution.}} \label{Polarization}
\end{center}
\end{figure}

Since the resonance  proposed here is obtained by fitting to the
existing data, a critical check would be to see whether additional
data would substantiate our interpretation. Accordingly, we also
calculate  the predictions of our model with and without the
inclusion of the proposed resonance for all the polarization
observables \cite{titov97}. In general, we find that the effects of the
resonance are substantial in many of the polarization observables {\cite{titov_polarization}}. We
show in Figs.~\ref{Polarization} our predictions for some of them
like the single polarization observables $\Sigma_x, T_y,$ in the
upper panel, and, in the lower panel, the double polarization
observables $C^{BT}_{yz}$ and $C^{BT}_{zx}$ at $E_\gamma=$ 2 GeV. It
is seen that the effects of the proposed resonance are huge in these
polarization observables. In the same figure, results that would be
obtained if the $3/2^+$ resonance determined in our best fitting is
adopted, are also shown by dash-dotted curve. We see that
measurements of these polarization observables would help to
resolve the question of the parity of the resonance.

\vspace{0.25cm}

From the $\phi-\omega$ mixing, one would expect that a resonance
 in $\phi N$ channel would  also appear in $\omega N$ channel. The only question
 is their relative decay strength.
 The conventional "minimal" parametrization relating $\phi NN^*$ and
 $\omega NN^*$ is
\begin{eqnarray}
  g_{\phi N N^*} = -\tan \Delta \theta_V
x_{\text{OZI}} g_{\omega N N^*}, \label{coupligrelation},
\end{eqnarray}
with $\Delta \theta_V \simeq 3.7^\circ$ corresponds to the
deviation from the ideal $\phi-\omega$ mixing angle.
Here, $x_{\text{OZI}}$ is called the OZI-evading parameter and the
larger value of $x_{\text{OZI}}$ would indicate  larger strangeness
content of the  resonance.

\begin{figure}[t]
\begin{center}
\vspace{0.5cm}

\includegraphics[width=0.5\linewidth,angle=0]{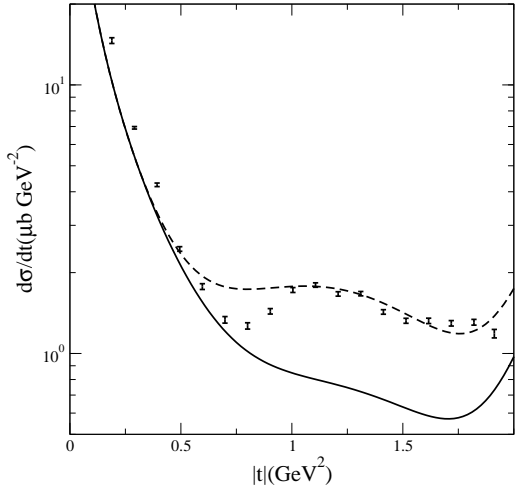}

\caption{Differential cross section of $\omega$ photoproduction as a
function of $|t|$ at $W = 2.105$ GeV. Solid and dashed lines
represent the model predictions of Ref. \cite{oh02} without and with
the addition of our resonance   with $x_{\text{OZI}} = 12$. Data are
from Ref. \cite{M_Williams}.} \label{omega}
\end{center}
\end{figure}

In order to study the effects of the resonance  postulated here in the
$\omega$ photoproduction, we adopt the study of Ref.~\cite{oh01}
which employs the nucleon resonances predicted by
Refs.~\cite{capstick1, capstick2}. In Fig.~\ref{omega}, it is seen
that the  prediction of their model for the $t$-dependence of differential cross section at
$W = 2.105$ GeV, given in solid lines, still exhibits substantial
discrepancy with  the most recent experimental data
\cite{M_Williams} for $|t| > 0.75$ GeV$^2$. By adding resonance
postulated here to the model of Ref. \cite{oh02} with $
x_{\text{OZI}}=12$, whose prediction is given in the dashed line in
Fig.~\ref{omega}, we see that the differential cross section at $W = 2.105$ GeV can be
reproduced with roughly the correct strength. The large value of
$x_{\text{OZI}}=12$ would imply that the resonance we propose here
contains a considerable amount of strangeness content.

In summary, we have explored the possibility of accounting for the
nonmonotonic behavior as observed by the LEPS collaboration at
energies close to threshold as a manifestation of a resonance. We
carry out calculations using a model with a nonresonant contribution which consists of
Pomeron plus $t$-channel $(\pi,\eta)$-exchange amplitudes, and a resonant contribution.
With resonance mass and width, and coupling constants as
parameters,  we perform a best fit to all the LEPS data at low
energies with possible assignments of $J = 1/2^\pm$ and $J =
3/2^\pm$.

We confirm that nonresonant contribution alone cannot describe the
nonmonotonic behavior of the forward differential cross section
near threshold and the $t$-dependence of the differential cross
section \cite{titov07}. We find that the addition of a resonance
with $J = 1/2$ of either positive or negative parity cannot explain
the local maximum at around $E_{\gamma}\sim 2.0$ GeV. However, with
an assignment of $J=3/2$, a nice agreement with most of the LEPS
data can be achieved.  We prefer the choice of $J=3/2^-$ as the best
fit to the data since its results are more stable with respect to changes in the low-energy Pomeron parameters. 
The obtained resonance mass and width are
$2.10\pm 0.03$ and $ 0.465\pm 0.141$ GeV, respectively.  The
resulting coupling constants give rise to a ratio of the helicity
amplitudes $A_{\frac 12}/A_{\frac 32}=1.16$, which differs from that
of the known $D_{13}(2080)$ in sign.


Furthermore, we find that the postulated resonance gives substantial
contribution to the polarization observables, which can also be used
to determine the parity of the resonance if it indeed exists.

The possible effects of this postulated resonance in the $\omega$
photoproduction are investigated by incorporating it within a recent calculation \cite{oh02} for this reaction. 
It turns out  that the addition of our resonance, with a choice of a large value of OZI-evading
parameter $x_{\textrm{OZI}}=12$, could indeed considerably improve
the agreement of the model prediction with the most recent data.
That would imply the resonance postulated here does contain
considerable amount of strangeness content.

There are a few caveats in our study. The first concerns the low-energy Pomeron parameters 
which are not presently very precisely
determined. If the postulated resonance contains considerable amount
of strangeness, then it could couple  strongly to, say, $K\Lambda$
channel. Question would then arise on how the coupled-channel
effects would modify the low-energy behavior of the nonresonant
amplitude employed in this investigation. This can be answered only
with a full coupled-channel calculations as carried out in Ref. \cite{ozaki09}. Another question is the validity of our
assumption to account for the local maximum with just one resonance. As
seen in the calculation of the effects of our postulated resonance,
some discrepancies with the recent data still persist after the
addition of this resonance. Accordingly, our study may have raised
more questions than it answers. Clearly, further studies, both
experimentally and theoretically, are needed on the $\phi$-meson
photoproduction at low energies.


\section*{Acknowledgments}
We would like to thank Profs. W. C. Chang, C. W. Kao, Atsushi Hosaka,
T.-S. H. Lee, Yongseok Oh, Sho Ozaki, and A. I. Titov, for useful
discussions and/or correspondences. This work is supported in part
by the National Science Council of Taiwan under grant
NSC98-2112-M002-006. We would also like to acknowledge the help from
National Taiwan University High-Performance Computing Center in
providing us with a fast and dependable computational environment
which is essential in completing this work.

\bibliographystyle{elsarticle-num}

\end{document}